\documentclass[nofootinbib,aps,prd,preprint,12pt]{revtex4}

\usepackage{amsmath,amsfonts}
\usepackage{feynarts}

\def\ket#1{\langle #1\rangle}
\def\P{\mathbb{P}}
\def\C{\mathbb{C}}
\def\Q{\mathbb{Q}}
\def\Z{\mathbb{Z}}

\begin{document}

\noindent
Brown-HET-1612 \hfill NSF-KITP-11-076

\vskip 1 cm 

\title{Symbols of One-Loop Integrals From Mixed Tate Motives}

\author{Marcus Spradlin}

\affiliation{Brown University, Providence, RI 02912, USA}
\affiliation{Kavli Institute for Theoretical Physics, University
of California, Santa Barbara, CA 93106, USA}

\author{Anastasia Volovich}

\affiliation{Brown University, Providence, RI 02912, USA}
\affiliation{Kavli Institute for Theoretical Physics, University
of California, Santa Barbara, CA 93106, USA}

\begin{abstract}

We use a result on mixed Tate
motives due to Goncharov~\cite{Goncharov:1996}
to show that the symbol of an arbitrary one-loop $2m$-gon integral
in $2m$ dimensions may be read off directly from its Feynman parameterization.
The algorithm proceeds via recursion in $m$ seeded by the well-known
box integrals in four dimensions.
As a simple application of this
method we write down the symbol of a three-mass hexagon integral in
six dimensions.
\end{abstract}

\maketitle

\section{Motivation}

Recent breathtaking advances in our understanding of the mathematical
structure of scattering
amplitudes\footnote{See
for example~\cite{RSV, ArkaniHamed:2010gh, Gaiotto:2011dt}
for reviews and recent progress.}
in maximally supersymmetric
Yang-Mills theory (SYM) have
so far had the most dramatic impact on
tree-level amplitudes or quantities (such as leading singularities or
the planar integrand~\cite{ArkaniHamed:2010kv})
which are completely determined in terms of
tree-level data.  The guiding principle behind these advances
is the observation that amplitudes exhibit simplicity which is completely
obscured by their traditional Feynman diagram expansions.

In contrast the problem of actually evaluating the integrals which
appear in multi-loop amplitudes remains extremely difficult in general
(see~\cite{Smirnov} for an introduction to modern methods).
However data is starting to accumulate which suggests that the
final results for amplitudes (or even individual integrals) can again be
far simpler than one might have guessed on the basis of results obtained
from more traditional integration approaches, such as Mellin-Barnes techniques.

In~\cite{Goncharov:2010jf} Goncharov, Vergu and the authors introduced to the
SYM literature a powerful mathematical tool first developed
in~\cite{Goncharov:2002,Goncharov:1994,Goncharov:1998},
called the symbol of a transcendental function,
for analyzing
functions of the type which appear in
all currently known SYM loop
amplitudes\footnote{See~\cite{Alday:2010jz,Gaiotto:2011dt,DelDuca:2011ne,Dixon:2011ng,DelDuca:2011jm,DelDuca:2011wh}
for recent applications of symbol technology.}.
The symbol serves as a sort of motivic
roadmap, encoding in a simple way all of the information about
a function's discontinuities without the need to introduce any
explicit representation of the function in terms of (generalized)
polylogarithm functions. 
Indeed one might reasonably
say that the symbol encapsulates all of the physically relevant
information that one may want to know about
any given integral or amplitude.
We view the symbol as stepping stone halfway between an integrand and
its integral, and it is our hope that it may
help to guide us through the
dark jungle of multi-loop amplitudes.

The simple formula
given in~\cite{Goncharov:2010jf} for the two-loop six-point
maximally helicity violating (MHV)
remainder function was made possible by applying motivic technology to the
17-page analytic expression which had been heroically evaluated
by Del Duca, Duhr and Smirnov~\cite{DelDuca:2009au,DelDuca:2010zg}.
It is clear that we are now
desperately in need of some technology which would allow one to
directly compute the symbol of any given integral without the need
for such heroic effort.  Since the planar integrand of SYM is a rational
form, and since the symbol of an integral involves only rational
functions, it is reasonable to suppose that there should exist an
algebraic operation which would allow one to simply write down the
symbol of any given integrand.  This would provide a concrete
realization of part of the ambitious program suggested
in~\cite{Goncharov:2002} (see in particular section 7)
which aims for a ``correspondence principle'' unifying scattering
amplitudes, their motivic avatars, and the combinatorics of
Feynman diagrams reflected in the coproduct formula.
There too the symbol is not the ultimate goal but
serves as just the simplest incarnation of
the amplitudes and their motivic structure.

Work towards unlocking the structure of general multi-loop
integrals
is underway, but given the recent flurry of
activity~\cite{DelDuca:2011ne,Dixon:2011ng,DelDuca:2011jm,DelDuca:2011wh} on
one-loop hexagon integrals in six dimensions we feel compelled to
focus in this brief note on the special case of the one-loop $2m$-gon integral
in $2m$ dimensions.  The motivic structure of this integral is sufficiently
simple that its symbol can be read off directly from its Feynman
parameterization by utilizing a 1996 theorem on mixed Tate motives
due to A.~Goncharov.  These higher dimensional integrals have several
applications in physics, for example they appear in
dimensionally-regulated one-loop
MHV amplitudes~\cite{Bern:1996ja,DelDuca:2009ac,Kniehl:2010aj}
and in certain cases they can be related to higher-loop four-dimensional
integrals~\cite{Drummond:2010cz,Anastasiou:2011zk}.

In section II we set up some notation.  The main result is presented
in section III, followed by several sample applications in section IV,
including a particular three-mass hexagon integral in six dimensions.
Note that in this note we work entirely at the level of symbols
and do not address the problem of choosing an explicit
representative for integrals in terms of (generalized) polylogarithm
functions.  It is our understanding that efforts to automate this final
step are under way~\cite{Duhr:2011zq}.

\section{Motivic Avatars of One-Loop Scalar Integrals}

In this note we consider the
one-loop $2m$-gon integral in $2m$ dimensions, specifically
\begin{equation}
\label{eq:Fzero}
F_m(y_i) = \Delta \int \frac{d^{2m} y}{(i \pi)^m} \prod_{i=1}^{2m}
\frac{1}{(y - y_i)^2}
\end{equation}
considered as a function of $2m$ points $y_1,\ldots,y_{2m}$
(with an implied cyclic ordering)
in $2m$-dimensional Minkowski space.
We have chosen to include in this definition the prefactor
\begin{equation}
\Delta = \sqrt{(-1)^m \det y_{ij}^2}, \qquad y_{ij} = y_i - y_j
\end{equation}
which serves three
intimately related purposes:
it ensures that $F_m(y_i)$ is invariant under dual conformal
transformations~\cite{Broadhurst:1993ib,Drummond:2006rz},
it normalizes the leading singularities of the integrand to $\pm 1$,
and it ensures that $F_m$
evaluates to a function of well-defined transcendentality $m$.
(We caution the reader that our $\Delta$ is close, but not exactly
equal, to the quantity called $\Delta$ in~\cite{DelDuca:2011ne,Dixon:2011ng}.)

Results for the $m=2$ integrals have of course been known since
antiquity (see for example~\cite{Bern:1993kr} for a comprehensive
modern
treatment),
while only very recently the first two cases of $m=3$ have
been studied:  the massless hexagon in~\cite{DelDuca:2011ne,Dixon:2011ng}
and the one-mass
hexagon in~\cite{DelDuca:2011jm}.
We remind the reader that
the integral~(\ref{eq:Fzero}) is said to be massless if
cyclically adjacent pairs are all null-separated, i.e. $y_{i,i+1}^2 = 0$
for all $i$.  More generally we say that the integral has $k$ masses
if $k$ pairs of cyclically adjacent points are null-separated.
For $m > 2$ the integral $F_m(y_i)$ always converges, while for
$m=2$ (the case of box integrals in four dimensions) it only
converges in the fully massive case.  We will return to the question
of divergences in detail below.

The $d^{2m} y$ integral is easily performed after introducing
Feynman parameters to combine the $2m$ propagators, which leads to
\begin{equation}
\label{eq:Fone}
F_m(y_i) = \Gamma(m) \Delta \int_0^\infty d^{2m} \alpha 
\frac{\delta(\alpha_1 - 1)}{\left(\sum_{i < j} y_{ij}^2\, \alpha_i \alpha_j\right)^m}\,.
\end{equation}
One may choose any nontrivial linear combination
of the $\alpha$'s inside the delta-function~\cite{Smirnov}
but our simple choice of
$\alpha_1$ makes it immediately manifest that~(\ref{eq:Fone}) is
best thought of as a projective
integral\footnote{We thank N.~Arkani-Hamed for
emphasizing this point.}.  Specifically~(\ref{eq:Fone}) is an
integral over $\C\P^{2m-1}$ with homogeneous
coordinates $[\alpha_1:\cdots:\alpha_{2m}]$ which has been
written in the patch $\alpha_1 = 1$.

Therefore, the one-loop $2m$-gon integral in $2m$ dimensions is
the $\C\P^{2m-1}$ period integral\footnote{See~\cite{Bogner:2007mn}
for a general discussion of the relation
between Feynman integrals and periods.}
\begin{equation}
\label{eq:Ftwo}
F_m(Q) = \Gamma(m)
\int D^{2m-1}W \frac{\Delta}{(\frac{1}{2} W \cdot Q \cdot W)^m}\,, \qquad
\Delta = \sqrt{(-1)^m\det Q}
\end{equation}
where $D^{2m-1} W$ is the standard holomorphic form
\begin{equation}
D^{2m-1} W = \epsilon^{i_1\cdots i_{2m}} w_{i_1} dw_{i_2} \wedge \cdots
\wedge dw_{i_{2m}}
\end{equation}
and where
we now use the notation $F_m(Q)$
to emphasize that $F_m$ should be thought of
as a function of a quadratic form $Q$,
whose matrix
entries are $Q_{ij} = (y_i - y_j)^2$ in our application.

Integrals of the type~(\ref{eq:Ftwo}) have been studied
in the context of mixed Tate motives, and belong to a class of
objects called Tate iterated integrals in~\cite{Goncharov:2009} (see
especially section 3).
In particular $F_m(Q)$ is a special
case of the more general period integrals $v(Q,M)$ studied
in~\cite{Goncharov:1996}.
These are integrals of the type~(\ref{eq:Ftwo}) involving any non-degenerate
quadric $Q$ and integrated over a cycle representing a generator
of $H_{2m-1}(\C\P^{2m-1},M)$ for an arbitrary
simplex $M$ in general position relative to $Q$.
The cohomology group $H^{2m-1}(\C\P^{2m-1}\backslash Q,M;\Q)$ has canonical
mixed Hodge-Tate structure which corresponds to a certain
mixed Tate motive~\cite{Goncharov:1996}.

In our particular application we shall only be interested
in quadratic forms $Q$ with vanishing diagonal entries
(since obviously $y_{ii}^2 = 0$ for all $i$).  Moreover,
the cycle of integration of interest in~(\ref{eq:Ftwo}), inherited
from the identification with~(\ref{eq:Fone}), involves
the particularly simple simplex $M$ whose $2m$ faces are
just the coordinate
hyperplanes $M_1,\ldots,M_{2m}$.
These two simplifications of our particular $F_m(Q)$'s compared to the
more general $v(Q,M)$'s studied in~\cite{Goncharov:1996}
are not of essential mathematical
importance
but greatly streamline the bookkeeping involved in applying the results
of that paper.

\section{An Auto-Motive Recursion for One-Loop Symbols}

We will now write a simple recursive
formula for the symbol $S_m(Q)$ of $F_m(Q)$.
The recursion expresses $S_m(Q)$ as a sum of $m(2m-1)$ terms
involving $S_{m-1}(Q_{\overline{ij}})$, where the notation
$Q_{\overline{ij}}$ denotes the $(2m-2)\times(2m-2)$ matrix obtained
by deleting the rows and columns $i$ and $j$ from the
$2m\times 2m$ matrix $Q$ (i.e., the quadric $Q_{\overline{ij}}$ is just
the intersection $Q \cap M_i \cap M_j$).
In terms of the Feynman graph corresponding to the
one-loop integral, going from $Q$ to $Q_{\overline{ij}}$
amounts to turning
the $2m$-gon into a $2m-2$-gon by deleting
the two propagators $(y - y_i)^2$ and $(y-y_j)^2$.
There are $m(2m-1)$ different ways of doing this, 
and each way leads to one term in the recursive sum.

To obtain the recursion we apply Theorem (4.10)
of~\cite{Goncharov:1996} (preprint version),
which for us becomes
\begin{equation}
\label{eq:dF}
d F_m(Q) = \frac{1}{2} \sum_{i < j}
F_{m-1}(Q_{\overline{ij}})
\ d \log R_{ij}\,,
\end{equation}
where
\begin{equation}
\label{eq:Rdef}
R_{ij} = \frac{Q^{-1}_{ij} + \sqrt{(Q^{-1}_{ij})^2 - Q^{-1}_{ii} Q^{-1}_{jj}}}
{Q^{-1}_{ij} - \sqrt{(Q^{-1}_{ij})^2 - Q^{-1}_{ii} Q^{-1}_{jj}}}\,.
\end{equation}
Note that no summation of indices is implied on the right-hand side,
which is expressed in terms of the $ii$, $ij$ and
$jj$ entries of the inverse matrix $Q^{-1}$.

Let us comment briefly on the geometric content of~(\ref{eq:Rdef}).
For each choice of $i$ and $j$, there are precisely two
codimension-1
hyperplanes
$H_1$, $H_2$ in $\C\P^{2m-1}$ which contain the intersection
$M_i \cap M_j$ and which are tangent to the quadric
$Q = \{ W \in \C\P^{2m-1} : \frac{1}{2} W \cdot Q \cdot W = 0 \}$.
So for each $i$ and $j$ we have four natural codimension-1 hyperplanes
$\{M_i,M_j,H_1,H_2\}$ all of which contain $M_i \cap M_j$ as a common
codimension-2 subspace, and the quantity $R_{ij}$ in~(\ref{eq:Rdef}) is
the cross-ratio naturally associated to this configuration.

It follows immediately from~(\ref{eq:dF}) that the symbol $S_m(Q)$ satisfies
the recursion
\begin{equation}
\label{eq:symbolrecursion}
S_m(Q) = \frac{1}{2} \sum_{i < j} S_{m-1}(Q_{\overline{ij}}) \otimes
R_{ij}\,.
\end{equation}
In what follows we shall only make use of~(\ref{eq:symbolrecursion}), though
of course~(\ref{eq:dF}) is a slightly stronger statement which realizes
a precise set of differential relations between the infinite class of
one-loop $2m$-gon integrals.
Various types of differential relations amongst multi-loop integrals
have long been studied; see~\cite{Drummond:2010cz,Dixon:2011ng}
for some recent
examples which also relate integrals in different dimensions.

\section{Sample Applications of the MB (Motivic Beauty) Technique}

We begin with a couple of almost trivial examples demonstrating
the application of the recursion~(\ref{eq:symbolrecursion}).
In general mathematical applications one may seed the recursion with
the result of the $\C\P^1$ integral
\begin{equation}
\label{eq:Fdegreeone}
F_1(Q) = \int D^1 W \frac{\Delta}{(\frac{1}{2} W \cdot Q \cdot W)}
= \log \frac{Q_{12} + \Delta}{Q_{12} - \Delta}, \qquad
\Delta = \sqrt{-\det Q}\,.
\end{equation}
However in physics applications, where we are always interested
in quadrics $Q$ with zero entries down the diagonal, this `bubble'
integral is too singular to be of use by itself.

Instead it is more effective to seed the recursion with box integrals
($m=2$).
As mentioned above these are divergent except in the fully massive case.
Nevertheless, as we discuss in detail below, it is relatively straightforward
to apply this recursion to arbitrary integrals by introducing if necessary a
regularization parameter $\epsilon$ which is ultimately taken to
zero.
The finiteness of the integral $F_m(Q)$ for $m>2$ is reflected in the
fact that the regularization parameter always drops completely out
of the symbol $S_m(Q)$ after summing together all contributions
to~(\ref{eq:symbolrecursion}).

Before proceeding let us make an important comment regarding notation.
In the previous section we defined the integral $F_m(y_i)$ for
an arbitrary collection of $2m$ points $y_i$ in Minkowski space, without
regard to whether or not any cyclically adjacent pairs were null
separated.  In the remainder of this paper we shall adopt notation more
conventional in the physics literature, whereby we consider a collection
of $n \ge 2m$ points which are all mutually null, i.e. $(x_i - x_{i+1})^2 = 0$
for all $i$, but only a subset of $2m$ of these $x_i$'s will be chosen
as the $2m$ arguments to the function $F_m(y_i)$.

\subsection{The Four-Mass Box Integral}

Let us start with the four-mass box integral, which corresponds to the
quadric
\begin{equation}
{\hbox{\lower 50.pt\hbox{
\begin{feynartspicture}(100,100)(1,1)
\FADiagram{}
\FAProp(13,7)(13,13)(0.,){/Straight}{0}
\FAProp(13,13)(7,13)(0.,){/Straight}{0}
\FAProp(7,13)(7,7)(0.,){/Straight}{0}
\FAProp(7,7)(13,7)(0.,){/Straight}{0}
\FAProp(13,13)(17.2426, 17.2426)(0.,){/Straight}{0}
\FAProp(13,7)(17.2426, 2.75736)(0.,){/Straight}{0}
\FAProp(7,7)(2.75736, 2.75736)(0.,){/Straight}{0}
\FAProp(7,13)(2.75736, 17.2426)(0.,){/Straight}{0}
\FALabel(16,10)[]{$x_j$}
\FALabel(10,16)[]{$x_i$}
\FALabel(4,10)[]{$x_l$}
\FALabel(10,4)[]{$x_k$}
\end{feynartspicture}
}}}
\qquad
Q = \left(\begin{matrix}
0 & x_{ij}^2 & x_{ik}^2 & x_{il}^2 \cr
x_{ij}^2 & 0 & x_{jk}^2 & x_{jl}^2 \cr
x_{ik}^2 & x_{jk}^2 & 0 & x_{kl}^2 \cr
x_{il}^2 & x_{jl}^2 & x_{kl}^2 & 0 \cr
\end{matrix}
\right)
\end{equation}
in $\C\P^3$.
The recursion~(\ref{eq:symbolrecursion}) expresses
the symbol of this integral
as a sum over six $\C\P^1$ integrals
of the type~\ref{eq:Fdegreeone}):
\begin{equation}
\label{eq:sixterms}
\begin{aligned}
S_2(Q) = \frac{1}{2} &\Big(
S_1(Q_{\overline{12}}) \otimes R_{12} +
S_1(Q_{\overline{13}}) \otimes R_{13} +
S_1(Q_{\overline{14}}) \otimes R_{14}
\cr
&+
S_1(Q_{\overline{23}}) \otimes R_{23} +
S_1(Q_{\overline{24}}) \otimes R_{24} +
S_1(Q_{\overline{34}}) \otimes R_{34} \Big)\,.
\end{aligned}
\end{equation}
Using~(\ref{eq:Fdegreeone}) we find for example
\begin{equation}
S_1(Q_{\overline{12}}) = \frac{x_{kl}^2 + \sqrt{x_{kl}^4}}{x_{kl}^2 - \sqrt{x_{kl}^4}}\,, \qquad
R_{12} =
\frac{
x_{il}^2 x_{jk}^2 + x_{ik}^2 x_{jl}^2 - x_{ij}^2 x_{kl}^2 - \Delta
}{
x_{il}^2 x_{jk}^2 + x_{ik}^2 x_{jl}^2 - x_{ij}^2 x_{kl}^2 + \Delta
}
\end{equation}
where $\Delta = \sqrt{+\det Q}$ in this case.

Evidently the symbol $S_1(Q_{\overline{12}})$ is singular,
but this can be regulated
by replacing all of the zero entries on the diagonal of $Q$ with $\epsilon$.
Then, at the level of the symbol, we can replace
\begin{equation}
\label{eq:regularization}
S_1(Q_{\overline{12}}(\epsilon)) = \frac{
x_{kl}^2 + \sqrt{x_{kl}^4 - \epsilon^2}}{x_{kl}^2 - \sqrt{x_{kl}^4 - \epsilon^2}}
= \frac{x_{kl}^4}{\epsilon^2} + {\cal O}(1) \quad \Longrightarrow
\quad
\frac{x_{kl}^4}{\epsilon^2}\,.
\end{equation}
Here we have freely dropped an overall factor of $4$
since such numerical constants drop out of
the symbol.
Note that we could have introduced the $\epsilon$ into $R_{12}$ as well but
there is no need as this quantity has a finite limit, shown above, as
$\epsilon \to 0$.

Adding up the six contributions~(\ref{eq:sixterms}) we find first of all
that $\epsilon$ completely drops out of the symbol (an important
consistency check), and that the result
can be assembled into the form
\begin{equation}
\label{eq:fourmass}
\begin{aligned}
S_2(Q) &=
x_{il}^2 x_{jk}^2 \otimes
\frac{
-x_{il}^2 x_{jk}^2 + x_{ik}^2 x_{jl}^2 + x_{ij}^2 x_{kl}^2 - \Delta
}{
-x_{il}^2 x_{jk}^2 + x_{ik}^2 x_{jl}^2 + x_{ij}^2 x_{kl}^2 + \Delta
}
\cr
&+ x_{ik}^2 x_{jl}^2 \otimes
\frac{
+x_{il}^2 x_{jk}^2 - x_{ik}^2 x_{jl}^2 + x_{ij}^2 x_{kl}^2 - \Delta
}{
+x_{il}^2 x_{jk}^2 - x_{ik}^2 x_{jl}^2 + x_{ij}^2 x_{kl}^2 + \Delta
}
\cr
&+ x_{ij}^2 x_{kl}^2 \otimes
\frac{
+x_{il}^2 x_{jk}^2 + x_{ik}^2 x_{jl}^2 - x_{ij}^2 x_{kl}^2 - \Delta
}{
+x_{il}^2 x_{jk}^2 + x_{ik}^2 x_{jl}^2 - x_{ij}^2 x_{kl}^2 + \Delta
}
\end{aligned}
\end{equation}
which of course agrees
precisely with the symbol of the well-known four-mass box
function
\begin{equation}
F_2 = {\rm Li}_2(x_+/x_-) - {\rm Li}_2\left(\frac{1-x_+}{1-x_-}\right)
+ {\rm Li}_2\left(\frac{1-1/x_+}{1-1/x_-}\right) - (x_+ \leftrightarrow x_-)
\end{equation}
in terms of
\begin{equation}
x_\pm =
\frac{1}{2}
\frac{x_{il}^2 x_{jk}^2 + x_{ik}^2 x_{jl}^2 - x_{ij}^2 x_{kl}^2 \pm \Delta}
{x_{il}^2 x_{jk}^2}\,.
\end{equation}

\subsection{Other Box Integrals in Four Dimensions}

{}From a mathematical point of view it is perhaps simplest to
restrict one's attention
to the generic case of the completely massive $2m$-gon
integral, whose symbol may now be written down directly
using the recursion~(\ref{eq:symbolrecursion}) 
seeded by~(\ref{eq:fourmass}) at $m=2$.

However in physics we are often interested in integrals with fewer masses.
The application of the recursion~(\ref{eq:symbolrecursion}) in these
cases will call into play various divergent box integrals which can
be regulated in a manner similar to~(\ref{eq:regularization}).
For example consider the three-mass box integral with
corresponding quadric
\begin{equation}
{\hbox{\lower 50.pt\hbox{
\begin{feynartspicture}(100,100)(1,1)
\FADiagram{}
\FAProp(13,7)(13,13)(0.,){/Straight}{0}
\FAProp(13,13)(7,13)(0.,){/Straight}{0}
\FAProp(7,13)(7,7)(0.,){/Straight}{0}
\FAProp(7,7)(13,7)(0.,){/Straight}{0}
\FAProp(13,13)(17.2426, 17.2426)(0.,){/Straight}{0}
\FAProp(13,7)(17.2426, 2.75736)(0.,){/Straight}{0}
\FAProp(7,7)(2.75736, 2.75736)(0.,){/Straight}{0}
\FAProp(7,13)(2.75736, 17.2426)(0.,){/Straight}{0}
\FALabel(16,10)[]{$x_j$}
\FALabel(10,16)[]{$x_i$}
\FALabel(4,10)[]{$x_{k+1}$}
\FALabel(10,4)[]{$x_k$}
\end{feynartspicture}
}}}
\qquad
Q = \left(\begin{matrix}
0 & x_{i,j}^2 & x_{i,k}^2 & x_{i,k+1}^2 \cr
x_{i,j}^2 & 0 & x_{j,k}^2 & x_{j,k+1}^2 \cr
x_{i,k}^2 & x_{j,k}^2 & 0 & 0 \cr
x_{i,k+1}^2 & x_{j,k+1}^2 & 0 & 0 \cr
\end{matrix}
\right)\,.
\end{equation}
Its symbol may be obtained by setting $x_{kl}^2 = \epsilon$
in~(\ref{eq:fourmass}) and keeping only the leading-order terms
for small $\epsilon$, which leads to
\begin{multline}
\label{eq:eighteen}
S_2(Q) =
\frac{1}{2}
\left(
\frac{\epsilon^2\ x_{i,j}^4}{x_{i,k}^2 x_{i,k+1}^2 x_{j,k}^2 x_{j,k+1}^2} \otimes \frac{ x_{i,k}^2 x_{j,k+1}^2 }{ x_{i,k+1}^2 x_{j,k}^2 }
+ \frac{ x_{i,k}^2 x_{j,k+1}^2 }{ x_{i,k+1}^2 x_{j,k}^2 } \otimes
\frac{\epsilon^2\ x_{i,j}^4}{x_{i,k}^2 x_{i,k+1}^2 x_{j,k}^2 x_{j,k+1}^2}
\right)
\cr
\qquad + \frac{ x_{i,k+1}^2 x_{j,k}^2 }{ x_{i,k}^2 x_{j,k+1}^2 }
\otimes
\frac{(x_{i,k}^2 x_{j,k+1}^2 - x_{i,k+1}^2 x_{j,k}^2)^2}
{x_{i,k}^2 x_{i,k+1}^2 x_{j,k}^2 x_{j,k+1}^2}\,.
\end{multline}
Results for all other box integrals may be obtained as further specializations
of this formula. For example the two-mass easy or two-mass hard box functions
can be obtained by setting $x_{ij}^2 = \epsilon$ or $x_{jk}^2 = \epsilon$
respectively.  Note that if we use~(\ref{eq:fourmass}) as our starting
point then it is not necessary to introduce an
$\epsilon$ for each of the diagonal entries of $Q$; in
writing~(\ref{eq:fourmass}) all of those divergences have already been 
properly eliminated.  We only need to introduce an $\epsilon$ for any
zero entry immediately adjacent to the diagonal (i.e. for any massless
vertex of the corresponding diagram).

Just like we saw in~(\ref{eq:fourmass}),
all dependence
on the regularization parameter $\epsilon$ will necessarily drop out when
individually divergent contributions are properly assembled
to produce the symbol of any of the finite functions $F_m$ for $m>2$.
In practice this cancellation
can serve as a useful check against calculational errors.

\subsection{A Three-Mass Hexagon in Six Dimensions}

As our first non-trivial\footnote{Actually in this particular case it
seems possible to generate the symbol directly from the
Feynman parameterization~(\ref{eq:Fone}) with the help of
Mathematica's {\tt Integrate} command;
in our experience it generally gets stuck starting at
degree 4 where the classical polylogarithms do not provide
a sufficiently large basis of functions.}
demonstration of the main result let us
write the symbol for the three-mass hexagon integral in six dimensions
corresponding to the quadric
\begin{equation}
\label{eq:threemass}
{\hbox{\lower 62.5pt\hbox{
\begin{feynartspicture}(125,125)(1,1)
\FADiagram{}
\FAProp(14.0000, 10.0000)(12.0000, 13.4641)(0.,){/Straight}{0}
\FAProp(12.0000, 13.4641)(8.00000, 13.4641)(0.,){/Straight}{0}
\FAProp(8.00000, 13.4641)(6.00000, 10.0000)(0.,){/Straight}{0}
\FAProp(6.00000, 10.0000)(8.00000, 6.53590)(0.,){/Straight}{0}
\FAProp(8.00000, 6.53590)(12.0000, 6.53590)(0.,){/Straight}{0}
\FAProp(12.0000, 6.53590)(14.0000, 10.0000)(0.,){/Straight}{0}
\FAProp(14.0000, 16.9282)(12.0000, 13.4641)(0.,){/Straight}{0}
\FAProp(6.00000, 16.9282)(8.00000, 13.4641)(0.,){/Straight}{0}
\FAProp(2.00000, 10.0000)(6.00000, 10.0000)(0.,){/Straight}{0}
\FAProp(6.00000, 3.07180)(8.00000, 6.53590)(0.,){/Straight}{0}
\FAProp(14.0000, 3.07180)(12.0000, 6.53590)(0.,){/Straight}{0}
\FAProp(18.0000, 10.0000)(14.0000, 10.0000)(0.,){/Straight}{0}
\FALabel(15.1962, 13.0000)[]{$x_2$}
\FALabel(10.0000, 16.0000)[]{$x_1$}
\FALabel(4.80385, 13.0000)[]{$x_8$}
\FALabel(4.80385, 7.00000)[]{$x_7$}
\FALabel(10.0000, 4.00000)[]{$x_5$}
\FALabel(15.1962, 7.00000)[]{$x_4$}
\end{feynartspicture}
}}}
\qquad
Q = \left(\begin{matrix}
0 & 0 & x_{14}^2 & x_{15}^2 & x_{17}^2 & x_{18}^2 \cr
0 & 0 & x_{24}^2 & x_{25}^2 & x_{27}^2 & x_{28}^2 \cr
x_{14}^2 & x_{24}^2 & 0 & 0 & x_{47}^2 & x_{48}^2 \cr
x_{15}^2 & x_{25}^2 & 0 & 0 & x_{57}^2 & x_{58}^2 \cr
x_{17}^2 & x_{27}^2 & x_{47}^2 & x_{57}^2 & 0 & 0 \cr
x_{18}^2 & x_{28}^2 & x_{48}^2 & x_{58}^2 & 0 & 0 \cr
\end{matrix}
\right)
\end{equation}
in $\C\P^5$.
Here we have chosen for convenience an integral with $n=9$ 
particles, which manifestly exhibits a $\Z_3$ symmetry.

We shall for convenience restrict the external
kinematics to four dimensions (i.e.~all of the $x_i$ lie within
a common four-dimensional subspace of six-dimensional Minkowski space).
This allows us to express everything in terms of momentum
twistors~\cite{Hodges:2009hk}; in particular $x_{ij}^2 \propto
\ket{i\ i{+}1\ j\ j{+}1}$ where the constant of proportionality
is irrelevant as it cancels out in all conformally invariant quantities.

We make this choice for purposes of notational simplicity only
but we emphasize that the result~(\ref{eq:symbolrecursion}) is of
course valid for completely general kinematics\footnote{Indeed we note that
with four-dimensional kinematics there does not exist
any quadric integral of the form~(\ref{eq:Ftwo}) 
for the octagon or higher since for $m>3$
the determinant
$\det y_{ij}^2$ is identically zero for any collection of 
$2m$ points $y_i$ which lie in a common four-dimensional subspace.}.
A special feature of the three-mass
hexagon~(\ref{eq:threemass}) is that no square roots of momentum
twistor invariants appear in its symbol.  This would not be true
if any two adjacent vertices of the hexagon were massive.

The recursion~(\ref{eq:symbolrecursion}) expresses the symbol
$S_3(Q)$ as a sum of 15 terms, but we can use the $\Z_3$ symmetry
of this configuration to cast the result into the form\footnote{A file
containing the complete symbol with all terms written out explicitly
is included with the arXiv submission.}
\begin{equation}
\label{eq:result}
\begin{aligned}
(1 + g + g^2)
\Bigg[
&S_2(Q_{\overline{12}}) \otimes \frac{\ket{2358} \ket{12\overline{5}\cap\overline{8}}}{\ket{1258} \ket{23\overline{5}\cap\overline{8}}}
\cr
+&S_2(Q_{\overline{13}}) \otimes \frac{\ket{1238} \ket{2568} \ket{5789} \ket{23\overline{5}\cap\overline{8}}}{\ket{2358} \ket{2789} \ket{4568} \ket{56\overline{2}\cap\overline{8}}}
\cr
+&S_2(Q_{\overline{14}}) \otimes \frac{\ket{2358} \ket{2789} \ket{4568} \ket{45\overline{2}\cap\overline{8}}}{\ket{1238} \ket{2458} \ket{5789} \ket{23\overline{5}\cap\overline{8}}}
\cr
+&S_2(Q_{\overline{16}}) \otimes \frac{\ket{2358} \ket{2456} \ket{5789} \ket{78\overline{2}\cap\overline{5}}}{\ket{1235} \ket{2578} \ket{4568} \ket{23\overline{5}\cap\overline{8}}}
\cr
+&S_2(Q_{\overline{24}}) \otimes \frac{\ket{1238} \ket{2458} \ket{5789} \ket{12\overline{5}\cap\overline{8}}}{\ket{1258} \ket{2789} \ket{4568} \ket{45\overline{2}\cap\overline{8}}}
\Bigg]
\end{aligned}
\end{equation}
where $g: i \to i + 3$ is the shift-by-three operator and we use
the notation
\begin{equation}
\ket{ij \overline{k}\cap \overline{l}} =
\ket{i\ k{-}1\ k\ k{+}1} \ket{j\ l{-}1\ l\ l{+}1}
- \ket{j\ k{-}1\ k\ k{+}1} \ket{i\ l{-}1\ l\ l{+}1}\,.
\end{equation}
Appearing in the first two entries of the symbol are a two-mass easy
box $Q_{\overline{12}}$ and some three-mass boxes
$Q_{\overline{13}}$, $Q_{\overline{14}}$, $Q_{\overline{16}}$ and
$Q_{\overline{24}}$, corresponding to the
degenerations of the hexagon:
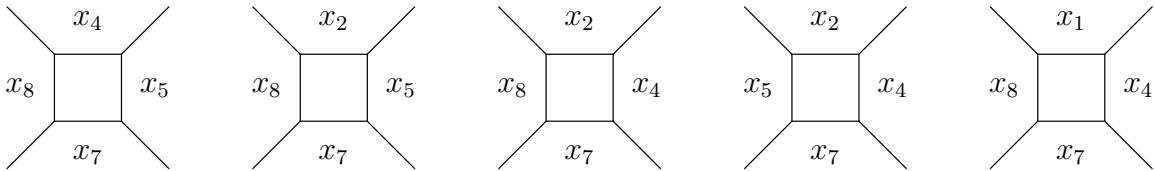
\begin{center}
\begin{feynartspicture}(465,93)(5,1)
\FADiagram{}
\FAProp(13,7)(13,13)(0.,){/Straight}{0}
\FAProp(13,13)(7,13)(0.,){/Straight}{0}
\FAProp(7,13)(7,7)(0.,){/Straight}{0}
\FAProp(7,7)(13,7)(0.,){/Straight}{0}
\FAProp(13,13)(17.2426, 17.2426)(0.,){/Straight}{0}
\FAProp(13,7)(17.2426, 2.75736)(0.,){/Straight}{0}
\FAProp(7,7)(2.75736, 2.75736)(0.,){/Straight}{0}
\FAProp(7,13)(2.75736, 17.2426)(0.,){/Straight}{0}
\FALabel(16,10)[]{$x_5$}
\FALabel(10,16)[]{$x_4$}
\FALabel(4,10)[]{$x_8$}
\FALabel(10,4)[]{$x_7$}
\FADiagram{}
\FAProp(13,7)(13,13)(0.,){/Straight}{0}
\FAProp(13,13)(7,13)(0.,){/Straight}{0}
\FAProp(7,13)(7,7)(0.,){/Straight}{0}
\FAProp(7,7)(13,7)(0.,){/Straight}{0}
\FAProp(13,13)(17.2426, 17.2426)(0.,){/Straight}{0}
\FAProp(13,7)(17.2426, 2.75736)(0.,){/Straight}{0}
\FAProp(7,7)(2.75736, 2.75736)(0.,){/Straight}{0}
\FAProp(7,13)(2.75736, 17.2426)(0.,){/Straight}{0}
\FALabel(16,10)[]{$x_5$}
\FALabel(10,16)[]{$x_2$}
\FALabel(4,10)[]{$x_8$}
\FALabel(10,4)[]{$x_7$}
\FADiagram{}
\FAProp(13,7)(13,13)(0.,){/Straight}{0}
\FAProp(13,13)(7,13)(0.,){/Straight}{0}
\FAProp(7,13)(7,7)(0.,){/Straight}{0}
\FAProp(7,7)(13,7)(0.,){/Straight}{0}
\FAProp(13,13)(17.2426, 17.2426)(0.,){/Straight}{0}
\FAProp(13,7)(17.2426, 2.75736)(0.,){/Straight}{0}
\FAProp(7,7)(2.75736, 2.75736)(0.,){/Straight}{0}
\FAProp(7,13)(2.75736, 17.2426)(0.,){/Straight}{0}
\FALabel(16,10)[]{$x_4$}
\FALabel(10,16)[]{$x_2$}
\FALabel(4,10)[]{$x_8$}
\FALabel(10,4)[]{$x_7$}
\FADiagram{}
\FAProp(13,7)(13,13)(0.,){/Straight}{0}
\FAProp(13,13)(7,13)(0.,){/Straight}{0}
\FAProp(7,13)(7,7)(0.,){/Straight}{0}
\FAProp(7,7)(13,7)(0.,){/Straight}{0}
\FAProp(13,13)(17.2426, 17.2426)(0.,){/Straight}{0}
\FAProp(13,7)(17.2426, 2.75736)(0.,){/Straight}{0}
\FAProp(7,7)(2.75736, 2.75736)(0.,){/Straight}{0}
\FAProp(7,13)(2.75736, 17.2426)(0.,){/Straight}{0}
\FALabel(16,10)[]{$x_4$}
\FALabel(10,16)[]{$x_2$}
\FALabel(4,10)[]{$x_5$}
\FALabel(10,4)[]{$x_7$}
\FADiagram{}
\FAProp(13,7)(13,13)(0.,){/Straight}{0}
\FAProp(13,13)(7,13)(0.,){/Straight}{0}
\FAProp(7,13)(7,7)(0.,){/Straight}{0}
\FAProp(7,7)(13,7)(0.,){/Straight}{0}
\FAProp(13,13)(17.2426, 17.2426)(0.,){/Straight}{0}
\FAProp(13,7)(17.2426, 2.75736)(0.,){/Straight}{0}
\FAProp(7,7)(2.75736, 2.75736)(0.,){/Straight}{0}
\FAProp(7,13)(2.75736, 17.2426)(0.,){/Straight}{0}
\FALabel(16,10)[]{$x_4$}
\FALabel(10,16)[]{$x_1$}
\FALabel(4,10)[]{$x_8$}
\FALabel(10,4)[]{$x_7$}
\end{feynartspicture}
\end{center}
Their $\epsilon$-dependent symbols may be read off from~(\ref{eq:eighteen}),
and it is a nontrivial check of the correctness of~(\ref{eq:result}) that
all of the $\epsilon$ dependence completely drops out at the level of
the symbol.
It is also straightforward to verify that in the limit
$x_{24}^2 \to 0$, $x_{57}^2 \to 0$ and $x_{18}^2 \to 0$ the
result~(\ref{eq:result})
reproduces the symbol of the massless hexagon computed
in~\cite{DelDuca:2011ne,Dixon:2011ng}.

In conclusion let us emphasize that in contrast to the old MB
method, the new MB (motivic beauty) technique does not require
the evaluation of any integrals.  The only even slightly
nontrivial step in obtaining~(\ref{eq:result}) was inverting
the matrix $Q$ shown in~(\ref{eq:threemass}) and simplifying
the $R_{ij}$'s from~(\ref{eq:Rdef}) to the final form presented.
Generalizations of this method to projective integrals of the
type which appear in higher loop integrals remain under
active investigation.

In this note we have worked entirely at the level of
symbols, without addressing the important problem of finding
explicit representations for integrals in terms of (generalized) polylogarithm functions.
This remains an interesting open problem, especially since
our experience with~\cite{Goncharov:2010jf} has shown that
analytic expressions for amplitudes can turn out to be much simpler
than one might have guessed from their symbols.

\section*{Acknowledgments}

We are grateful to S.~Caron-Huot, J.~Henn, D.~Skinner, C.~Vergu
and especially N.~Arkani-Hamed and A.~Goncharov for useful
discussions, correspondence, and encouragement.
This work was supported in part by the
Department of Energy under contract DE-FG02-91ER40688 Task J OJI (MS)
and Task A (AV), and the National Science Foundation under Grant
Nos.~PHY05-51164,
PECASE PHY-0643150 (AV) and ADVANCE 0548311 (AV).

\end{document}